\documentclass[12pt]{iopart}
\usepackage{graphicx}% Include figure files
\usepackage{epstopdf}% Convert eps to pdf
\begin{document}

\title[]{Shear Viscosity in a Perturbative Quark-Gluon-Plasma}

\author{John Fuini III$^{1,2}$\footnote{current address: Department of Physics, University of 
Washington, Seattle, WA 98195-1560}, Nasser S. Demir$^2$ Dinesh K. Srivastava$^3$ and Steffen A. Bass$^2$}

\address{$^1$ The University of Texas at San Antonio, San Antonio, Texas 78249-1644, USA}
\address{$^2$ Department of Physics, Duke University,
             Durham, North Carolina 27708-0305, USA}
\address{$^3$ Variable Energy Cyclotron Centre,1/AF Bidhan Nagar, Kolkata 700 064, India}

\ead{john.fuini@gmail.com}

\begin{abstract}
Among the key features of hot and dense QCD matter produced in ultra-relativistic heavy-ion
collisions at RHIC is its very low shear viscosity, indicative of the properties of a near-ideal fluid, and a large opacity demonstrated by jet energy loss measurements.  In this work, we utilize a microscopic transport model based on the Boltzmann equation with quark and gluon degrees of freedom and cross sections calculated from perturbative Quantum Chromodynamics to simulate an ideal Quark-Gluon-Plasma
in full thermal and chemical equilibrium.
We then use the Kubo formalism to calculate the shear viscosity to entropy density ratio of the medium 
as a function of temperature and system composition. One of our key results is that the shear
viscosity over entropy-density ratio $\eta/s$ becomes invariant to the chemical composition
of the system when plotted as a function of energy-density instead of temperature.
\end{abstract}

\submitto{\JPG}
\maketitle

\section{Introduction}
Ultrarelativistic heavy ion collisions at the Relativistic Heavy Ion Collider (RHIC) are thought to have produced a Quark Gluon Plasma (QGP) with the characteristics of a near ideal fluid\cite{Adcox:2004mh,Back:2004je,Arsene:2004fa,Adams:2005dq}. One of the most important current challenges
in QGP research is to quantify the transport coefficients of this novel state of matter.  Recently, attention in the field has been primarily focused on the shear viscosity to entropy density ratio $\eta/s$.  Calculations utilizing certain strongly coupled supersymmetric gauge theories with gravity duals \cite{Policastro:2001yc} postulate a lower bound of $\eta_{\rm min} = s/4\pi$ for this quantity, often referred to as the KSS bound \cite{Kovtun:2004de}.  Relativistic viscous hydrodynamic calculations have confirmed very low values of $\eta/s$ in order to reproduce the RHIC elliptic flow ($v_2$) data \cite{Song:2007fn,Romatschke:2007mq,Luzum:2008cw}.  However, current calculations assume a fixed value of 
$\eta/s$ throughout the entire evolution of the system and neglect its temperature dependence,
which anyhow cannot be obtained from the AdS/QCD calculations.

It has been argued on very general grounds that $\eta/s$ should exhibit a minimum at
the phase-transition and should rise for decreasing temperature in the hadronic phase
and for increasing temperature in the deconfined phase \cite{Csernai:2006zz}.
Recent calculations of $\eta/s$ using microscopic transport theory in the hadronic sector
have confirmed this expectation and have provided quantitative guidance with respect to the temperature dependence of $\eta/s$ in and out of chemical equilibrium \cite{Demir:2008tr}. 
At temperatures above $T_C$ there remains a large uncertainty regarding the value 
and temperature-dependence of $\eta/s$:
in the limit of very high temperatures, the interactions between the constituents of QCD matter 
should be calculable with perturbative methods \cite{Arnold:2003zc}, whereas in the strongly interacting
regime near $T_C$ $\eta/s$ could be close to the KSS bound. It is our goal in this paper to
explore the temperature dependence of QCD matter for temperatures above $T_C$
at which quasi-particle degrees of freedom become viable. For our calculation we will
use a microscopic transport model, the BMS implementation of the Parton Cascade Model (PCM) 
\cite{Bass:2002fh},
which has the advantage that it can describe both, out of equilibrium QCD matter created in 
collisions of ultra-relativistic heavy-ions, as well as equilibrated QCD matter at
a fixed temperature. For the calculation of the shear viscosity we shall use the Kubo
formalism utilized in our previous calculation of $\eta$ and $\eta/s$ in the hadronic 
sector \cite{Demir:2008tr}.

\section{The Parton Cascade Model}
The Parton Cascade Model (PCM) \cite{Geiger:1991nj,Geiger:1997pf} is a microscopic transport model which is used to simulate the time evolution of a system of quarks and gluons utilizing the Boltzmann equation. Here, we use the
BMS implementation, which has been applied successfully to the calculation of direct photon production  \cite{Bass:2002pm,Renk:2005yg}  and baryon stopping \cite{Bass:2002vm} at RHIC. PCM implementations
by other groups have been used to address collective flow, parton energy-loss as well as
multi-particle effects on transport coefficients 
\cite{Molnar:2000jh,Molnar:2004yh,Xu:2004mz,Xu:2007ns,El:2008yy,Fochler:2008ts,Xu:2008av,Uphoff:2010fz}.
Our calculation focuses on an ideal Quark-Gluon-Plasma, i.e. a gas of
$u, d$ and $s$ quarks and anti-quarks as well as gluons at fixed temperature $T$ 
in full thermal and chemical equilibrium. In addition, we also study a one-component 
gluon plasma in thermal and chemical equilibrium. For our transport calculation we
define a box with periodic boundary conditions (to simulate infinite matter) and 
sample thermal quark and gluon distribution functions to generate an ensemble of
particles at a given temperature and zero chemical potential.

The time evolution of our system is described by a  Boltzmann transport equation:

\begin{equation}
p^\mu \frac{\partial}{\partial x^\mu} F_k(x,\vec p) =  \sum\limits_{{\rm processes:}~i}{\cal C}_i[F]
\label{eq03}
\end{equation}
where the collision terms ${\cal C}_i$ are nonlinear functionals
of the phase-space distribution function. Although the collision
term, in principle, includes factors encoding the Bose-Einstein 
or Fermi-Dirac statistics of the partons, we neglect those effects
here.

The collision integrals have the form:
\begin{equation}
\label{ceq1}
{\cal C}_i[F] = \frac{(2 \pi)^4}{2 S_i} \cdot
\int  \prod\limits_j {\rm d}\Gamma_j \, | {\cal M}_i |^2 
       \, \delta^4(P_{\rm in} - P_{\rm out}) \, 
          D(F_k(x, \vec p)) 
\end{equation}
with
\begin{equation}
D(F_k(x,\vec p)) \,=\, 
\prod\limits_{j={\rm out}} F_j(x,\vec p) \, - \, F_k(x,\vec p)
\prod\limits_{j={\rm in} \atop {j\ne k}} F_j(x,\vec p) \quad
\end{equation}
and
\begin{equation}
\prod\limits_j {\rm d}\Gamma_j = \prod\limits_{{j \ne k} \atop {\rm  in,out}} 
        \frac{{\rm d}^3 p_j}{(2\pi)^3\,(2p^0_j)} 
\quad.   
\end{equation}
$S_i$ is a statistical factor defined as
$S_i \,=\, \prod\limits_{j \ne k} K_a^{\rm in}!\, K_a^{\rm out}!$
with $K_a^{\rm in,out}$ identical partons of species $a$ in the initial
or final state of the process $i$, excluding the $k$th parton.

The matrix elements for the full Quark-Gluon-Plasma calculation
 $| {\cal M} |^2$ account for the following 
processes (note that all $t$ and $u$-channel processes are included as well):

%feynman diagrams
\begin{figure}[htbp]
\begin{center}
	\includegraphics[width=0.20\textwidth]{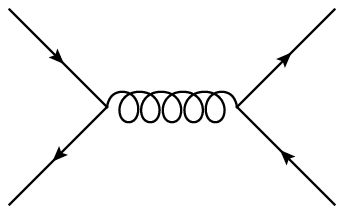}\hfill
	\includegraphics[width=0.20\textwidth]{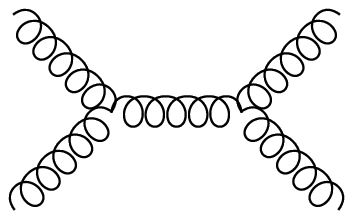}\hfill
	\includegraphics[width=0.20\textwidth]{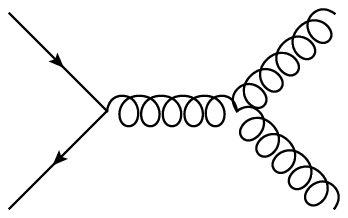}\hfill
	\includegraphics[width=19.5mm, height=19.5mm]{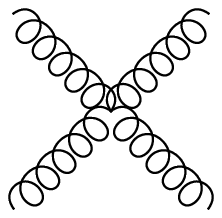}
\end{center}
\end{figure}

The 
corresponding scattering cross sections are expressed in terms
of spin- and color-averaged amplitudes $|{\cal M}|^2$:
\begin{equation}
\label{dsigmadt}
\left( \frac{{\rm d}\hat \sigma}
     {{\rm d} Q^2}\right)_{ab\to cd} \,=\, \frac{1}{16 \pi \hat s^2}
        \,\langle |{\cal M}|^2 \rangle
\end{equation}

In the case of a pure Gluon Plasma (GP), we use the following differential
cross section in order to facilitate comparisons with PCM implementations
by other groups \cite{Molnar:2000jh,Molnar:2004yh,Xu:2004mz,Xu:2007ns,El:2008yy}:
\begin{equation}
{{d\sigma^{gg\rightarrow gg}}\over{dQ^2 }} \,=\, 
{2\pi\alpha_s^2}  {9 \over 4} {1 \over {(Q^s + m_D^2 )^2}},
\label{diff_cs_gg_gg}
\end{equation}

For the transport calculation we also need the total cross section 
as a function of $\hat s$ which can be obtained from (\ref{dsigmadt}):
\begin{equation}
\label{sigmatot}
\hat \sigma_{ab}(\hat s) \,=\, 
\sum\limits_{c,d} \, \int\limits_{0}^{\hat s}
        \left( \frac{{\rm d}\hat \sigma }{{\rm d} Q^2}
        \right)_{ab\to cd} {\rm d} Q^2 \quad .
\end{equation}

The amplitudes for the above processes have been calculated in refs. 
\cite{Cutler:1977qm,Combridge:1977dm} for massless quarks. Note, however, that for the kinematics
of the system, the light quark masses are treated explicitly.
We regularize the IR divergence of the cross sections with a temperature dependent Debye mass $m_D$.
We shall use two different expressions for $m_D$ -- the first one is a Debye-mass for particles
obeying Boltzmann statistics which has been used in \cite{Xu:2004mz,El:2008yy,Fochler:2008ts,Xu:2008av,Uphoff:2010fz} and which we utilize to allow our results to be compared to these calculations:
\begin{equation}
m_D(T) = \sqrt{\frac{24}{\pi}\alpha_s T^2} \quad,
\end{equation}
and the second one is the standard Debye mass used in pQCD calculations for systems of 
quarks and gluons:
\begin{equation}
m_D(T) = gT \sqrt{(2N_c + N_f)/6} 
\end{equation}
The first $m_D$ parametrization we shall refer to as Boltzmann-$m_D$ whereas the second 
parametrization we shall refer to as regular $m_D$.
In both cases the coupling constant is defined as 
\begin{equation}
\alpha_s = \frac{g^2}{4\pi}
\end{equation}
and can either be chosen as a constant parameter, or to have the following
temperature dependence \cite{Csernai:2006zz}:
\begin{equation}
\frac{1}{g^2} = \frac{9}{8\pi^2}\ln\left(\frac{T}{\Lambda_T}\right) + \frac{4}{9\pi^2}\ln\left(2\ln\left(\frac{T}{\Lambda_T}\right)\right)
\end{equation}
with $\Lambda_T = 30$~MeV.

\section{Methodology}

\subsection{Shear-viscosity}

In order to calculate the shear-viscosity of our system, we employ the 
Kubo-formalism \cite{Kubo:1957mj,Kubo-formalism}. The methodology for 
applying the Kubo-formalism to infinite QCD matter modeled by microscopic
transport theory was developed in \cite{Muronga:2003tb} and has already been successfully
applied to calculating the shear viscosity for a hadron gas \cite{Muronga:2003tb,Demir:2008tr}.
The momenta of all the partons in our system 
are tracked over the course of 20 time-steps of 0.5 fm/c.  Knowing the momenta $p$ of all partons
in the system, the
discretized stress-energy tensor of our system can be calculated at each timestep:

%stress energy tensor
%stress energy tensor - discrete
\begin{equation}
\pi^{\mu\nu} = \int{d^3p}  \frac{p^{\mu}p^{\nu}}{p^0}f(x,p) \Rightarrow \frac{1}{V} \sum_{i=1}^N\frac{p^x(i)p^y(i)}{p^0}
\end{equation}

The stress-energy tensor now allows for the construction of the stress energy tensor correlator
used in the Kubo formalism \cite{Kubo:1957mj,Kubo-formalism} to calculate the shear viscosity via

%kubo eta
\begin{equation}
\eta=\frac{1}{T}\int{d^3r}\int_0^\infty dt \langle\pi^{xy}(0,0)\pi^{xy}(\vec{r},t)\rangle_{equil}
\end{equation}
where the average is taken over many events and $T$ is the temperature of the system.
A selection of correlators for different temperatures are shown in figure~1. Based
on the observed exponential decay of the correlators, we fit these with the following
analytic function: 
\begin{equation}
\langle\pi^{xy}(0,0)\pi^{xy}(\vec{r},t)\rangle \sim e^{-\frac{t}{\tau_R}}
\end{equation}
with the parameter $\tau_R$ called the relaxation time of the system. Inserting this function
into the expression for $\eta$ yields:
\begin{equation}
\eta = \frac{V}{T}\langle\pi^{xy}(0,0)\pi^{xy}(0,0)\rangle\tau_R
\end{equation}
with the volume $V$ and temperature $T$ being input parameters of our calculation and $\tau_R$ and 
$\langle\pi^{xy}(0,0)\pi^{xy}(0,0)\rangle$ determined by our exponential fit (see figure~1).

%CORRELATORS ANSANTZ
\begin{figure}[t]
\centerline{\includegraphics[width=0.55\textwidth]{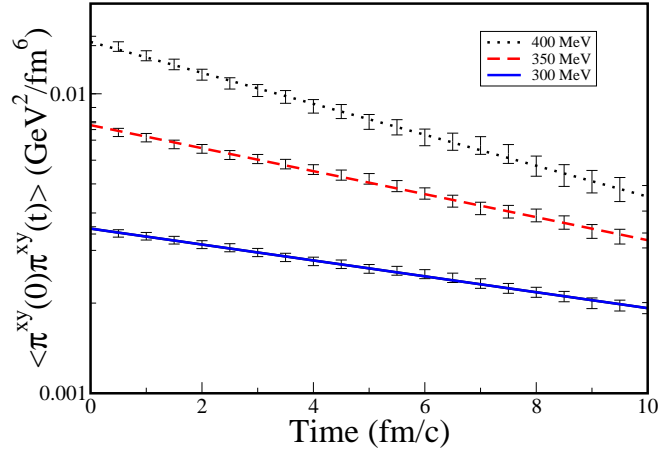}}
 \caption{Kubo correlators for a system of gluons at various temperatures. The correlators
 show a clear exponential decay.}
\end{figure}

\subsection{Entropy} %Why is this italicized!?

The entropy-density of our system is being calculated via the Gibbs entropy density:
\begin{equation}
s = \frac{\epsilon + P + \mu_B\rho_B}{T}
\end{equation}
Since the system is initialaized with net baryon density, $\rho = 0$, the pressure  $P$ and
energy-density $\epsilon$ are the only quantities which need to be extracted from 
our simulation according to 
\begin{equation} 
P = \frac{1}{3V}\sum_{i=1}^N \frac{\left|\vec{p}(i)\right|^2}{p^0(i)}
\qquad \mbox{and} \quad \epsilon = \frac{1}{V}\sum_{i=1}^Np^0(i) \quad.
\end{equation}
The left frame of figure~2 shows the resulting entropy-density for
a Gluon Plasma (GP) as well as a Quark-Gluon-Plasma (QGP) containing gluons
and three light quark flavors.

We can compare the entropy density directly obtained from our calculation  to 
the Stefan Boltzmann entropy density given by:
\begin{equation}
s = \frac{2\pi^2}{45}\nu(T)T^3
\end{equation}
where $\nu(T) = N_b + \frac{7}{8}N_f$ and $N_b$ and $N_f$ are the bosonic and fermionic degrees of freedom. The results of the comparison are shown in the left frame of and figure~2 and show
good agreement between the Stefan Boltzmann entropy and our system entropy calculated via 
the Gibbs relation.

\section{Results and Discussion}

We will present our results predominantly for two modes of calculation: the first one,
denoted by GP, is for a gluon plasma with a regular Deybe screening mass and temperature
dependent coupling. The second mode, denoted by QGP, uses the same temperature-dependent 
parametrizations for $m_D$ and the coupling constant, but is for a quark-gluon plasma with
three light quark flavors. We consider the QGP mode to be the most realistic mode of 
calculation presented here.

%ETS OVER S FINAL RESULTS
\begin{figure}[t]
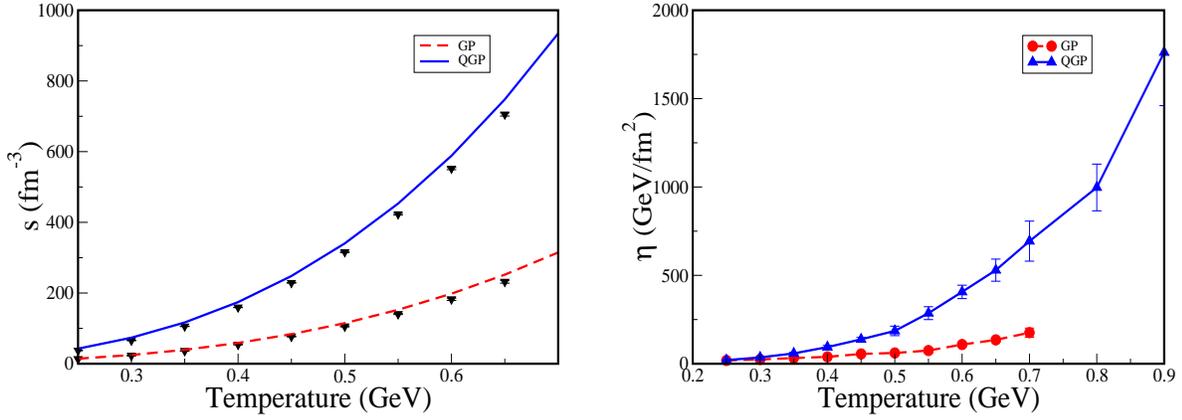

	\centering
		\includegraphics[width=0.470\textwidth, height = 55mm]{EntrvsT.eps}\hfill
		\includegraphics[width=0.470\textwidth, height = 55mm]{EtavsTnew.eps}
	\label{fig:EtaoverSfinal}
	\caption{Left: entropy as a function of temperature for a Gluon-Plasma and a three flavor
	Quark-Gluon-Plasma. The symbols denote the entropy of the system evaluated using
	the Gibbs formula and the lines represent the Stefan Boltzmann expression for the 
	entropy density with $\nu(T) = 16$, and Quark-Gluon-Plasma, $\nu(T) = 47.5$  
	Right: shear-viscosity $\eta$ as a function of temperature for the Gluon Plasma (GP) and
	the Quark-Gluon-Plasma (QGP).   }
\end{figure}

The right frame of Figure~2 displays the results of our calculation for the shear viscosity
for a system of gluons as well as a system of gluons and three light quark flavors. Both,
the shear viscosity, as well as the entropy density depicted in the left frame of figure~2,
rise strongly as a function of increasing temperature. It is interesting to note that the
QGP shows a significantly higher shear-viscosity than the GP for a given temperature,
which seems counter-intuitive given the larger particle density of the QGP, but is probably
due to the smaller interaction cross-sections among the quarks of the system.

Having calculated both, the shear-viscosity as well as the entropy-density of our system, we can
now turn to the ratio $\eta/s$, made famous by the KSS bound: the left frame of figure~3 shows
$\eta/s$ as a function of temperature for the GP (full circles) and QGP (full triangles) case, 
compared to an analytic HTL calculation of a three-flavor QGP \cite{Arnold:2003zc} (solid line).
The QGP calculation of $\eta/s$ shows a monotonous rise as a function of temperature with a slope
very similar to that of the HTL calculation. The differences in absolute value between the two
can be understood by considering the NLL corrections present in the HTL calculation into account,
which are missing in our simulation of QCD matter. The Gluon Plasma exhibits an upturn in 
$\eta/s$ for temperatures below 500~MeV -- we attribute this unexpected rise towards lower
temperatures to a breakdown in the perturbative
approximations present in our calculations, emphasized by taking the ratio of two quantities,
which both, when calculated and viewed separately (see figure~2), seem to exhibit a smooth behavior
in the low temperature domain below 500~MeV. 

\begin{figure}[t]
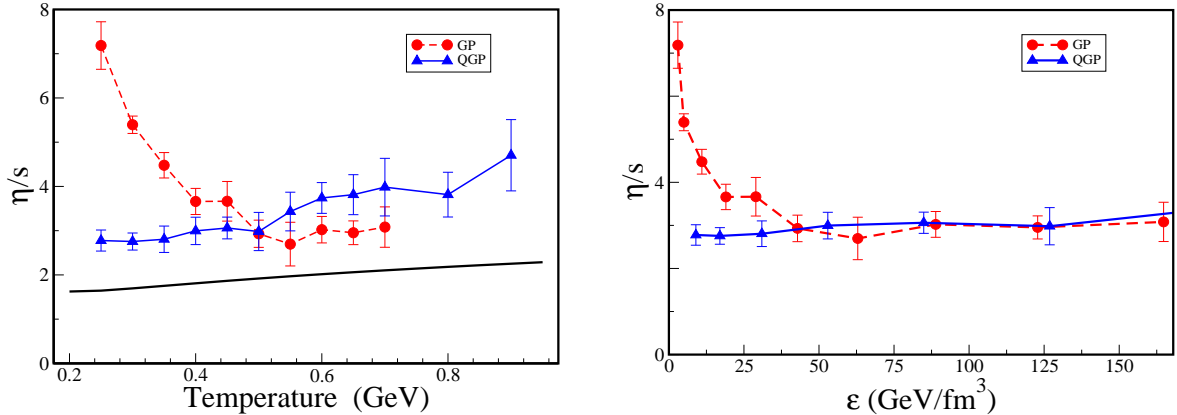

	\centering
		\includegraphics[width=0.470\textwidth, height = 55mm]{EtaoverSfinal.eps}\hfill
		\includegraphics[width=0.470\textwidth, height = 55mm]{EtaoverSfinalEpsilon.eps}
	\label{fig:alphasvariationetaoversZOOM}
	\caption{Left: $\eta/s$ as a function of temperature for a Gluon-Plasma (full circles) and Quark-Gluon-Plasma (full triangles), compared to a HTL calculation (solid line) \cite{Arnold:2003zc}. 
	Right: same result, but plotted vs. energy density.}
\end{figure}

Comparing $\eta/s$ of a GP and a QGP at the same temperature may be misleading due to the 
significantly larger parton density present in a QGP. Therefore in the right frame of figure~3
we compare $\eta/s$ for the two scenarios at equivalent energy-density and find for energy-densities
above 35 GeV/fm$^3$ excellent agreement between the two systems. This is of particular relevance
since the flavor composition of the deconfined QCD matter created in ultra-relativistic 
heavy-ion collisions is by no means fully established and may change strongly as a function
of time -- from a gluon-dominated system being created by the decay of a Color-Glass-Condensate
to a QGP in full thermal and chemical equilibrium as the system progresses in its hydrodynamic
expansion. Our result indicates that $\eta/s$, a quantity which controls the hydrodynamic evolution
of the system, should be fairly robust with respect to its flavor composition when taken
as a function of energy-density instead of temperature.

In the left frame of figure~4 we study the effects of our different Debye-mass parametrizations
and the temperature-dependence on $\eta/s$. For this purpose we calcuate $\eta/s$ for a gas
of gluons with a Boltzmann Debye-mass at fixed coupling of $\alpha_s=0.3$ (i.e. the same system as e.g.
in \cite{Xu:2004mz,Xu:2007ns,El:2008yy,Fochler:2008ts,Xu:2008av}) and compare this calculation to one
with our default parametrization (taken with $N_c=3$ and $N_f=0$). We find the effect of
the different  $m_D$ parametrizations to be small, on the 10\% to 15\% level, with the Boltzmann Debye-mass
giving systematically smaller values of $\eta/s$. If we now replace the fixed coupling with
a temperature dependent coupling constant, the value of $\eta/s$ increases roughly by a factor of $2$.

\begin{figure}[t]
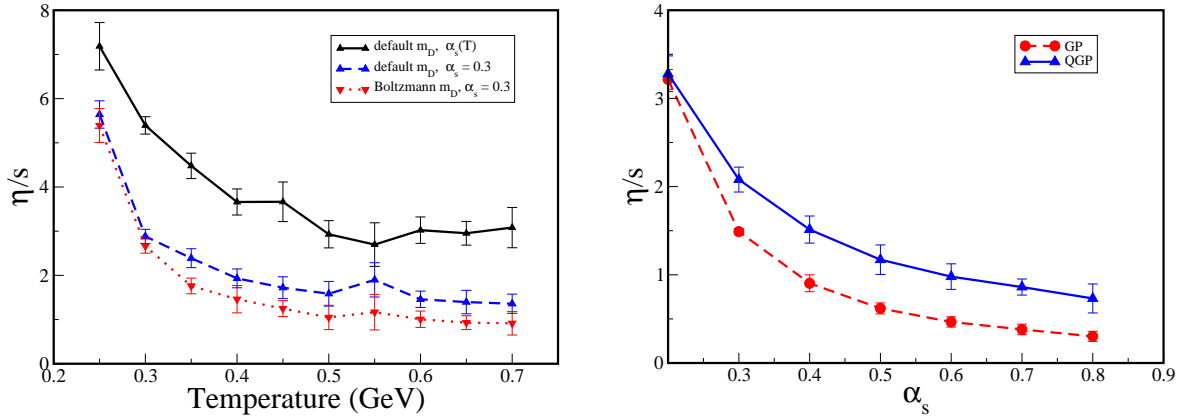

	\centering
		\includegraphics[width=0.470\textwidth, height = 55mm]{GOetaoverscomparison.eps}\hfill
		\includegraphics[width=0.470\textwidth, height = 55mm]{alphasvariationetaoversZOOM.eps}
	\label{fig:alphasvariationetaoversZOOM}
	\caption{Left: effect of different Debye mass parametrizations and temperature-dependent
	coupling on $\eta/s$. Right: Effect of coupling constant on $\eta/s$.}
\end{figure}

So far we have restricted our investigation to a purely perturbative partonic system with
the respective temperature-dependent screening masses and coupling strengths. However, there
are strong indications that the medium created in ultra-relativistic heavy-ion collisions
is non-perturbative in nature -- at the very least within the temperature range from $T_C$ to
approximately $3-4 T_C$, which is covered by our calculations. One method to explore the behavior
of $\eta/s$ at stronger coupling is to treat the coupling constant in our calculations as
a free parameter and then study $\eta/s$ at fixed temperature as a function of the 
coupling constant. The right frame of figure~4 shows $\eta/s$ as a function of coupling strength
for the gluon plasma and the quark-gluon-plasma. In the strong coupling limit, in particular
for the gluon plasma, values of $\eta/s \ll 1$ can be obtained, yielding results compatible
with a fluid-dynamical analysis of RHIC data \cite{Luzum:2008cw}. Similar results have been obtained by \cite{Molnar:2004yh},
who directly increased the gluon-gluon scattering cross-section, which is equivalent to an
increase in the coupling constant or by directly studying $\eta/s$ as a function of
the coupling constant \cite{Xu:2007ns}. One should note, however, that treating the coupling constant
as a free parameter yields a medium with inconsistent characteristics, since the particle 
density and screening mass are still controlled by the temperature, whereas the coupling is not.
Also, for large couplings $\alpha_s \simeq 1$ the perturbative assumptions underlying the PCM
are not valid anymore.

\begin{figure}[t]
	\centering
		\includegraphics[width=0.470\textwidth, height = 55mm]{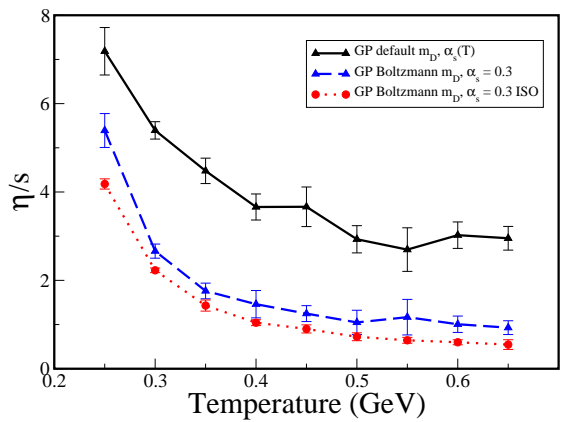}
	\label{fig:ISOcomparison}
	\caption{Coupling constant and angular distribution dependence of $\eta/s$ for a gluon plasma:
	The choice of a fixed coupling constant affects $\eta/s$ to a far larger degree than the
	transition from the regular angular distribution to isotropic scattering.}
\end{figure}

The final question we wish to address is the effect the angular distribution of the scattering
partons , i.e. the exact form of the differential matrix-element, has on $\eta/s$:
figure~5 shows a comparison between our default gluon plasma calculation
compared with results at fixed coupling and regular matrix
elements as well as with the same matrix elements using an isotropic angular distribution. 
The main effect, a reduction of $\eta/s$ by a factor of 3, comes from the choice
of a fixed coupling constant vs. a temperature dependent one. 
Changing from forward-backward peaked to isotropic scattering provides an additional 10\% - 20\% 
effect, but not a dramatic reduction in $\eta/s$.
This finding is
of particular interest in the context of work done by \cite{Xu:2004mz,Xu:2007ns,El:2008yy,Fochler:2008ts,Xu:2008av}, employing a PCM including
radiative corrections (i.e. $2 \to 3$ and $3 \to 2$ processes) and an implementation 
of the LPM effect, which manifests itself in a near-isotropic angular distribution for the 
third particle in the outgoing channel. It has been shown that the isotropic emission of the
radiated particle in this implementation of the LPM effect
plays an important role for the thermalization of the system.
Our results utilizing solely $2 \to 2$ scattering processes 
indicate that it is most likely a combination of multiple effects -- choice of fixed coupling
constant, the isotropic angular distribution of the LPM implementation, and the inclusion of
radiative corrections,
which is responsible for the low viscosity findings of these calculations \cite{Xu:2007ns,El:2008yy}.
 
\section{Summary}
We have utilized the parton cascade model to simulate a perturbative quark-gluon-plasma in
full thermal equilibrium and have extracted its shear-viscosity as well as the shear-viscosity
over entropy-density ratio as a function of temperature. We find that our results depend significantly
on the details of the calculation, i.e. choice of coupling constant and parametrization of the
Debye screening mass as well as the degrees of freedom (gluon vs. quark-gluon plasma).
One of our key results is that the shear
viscosity over entropy-density ratio $\eta/s$ becomes invariant to the chemical composition
of the system when plotted as a function of energy-density instead of temperature.
The values we obtain for $\eta/s$ are higher than those expected for the near ideal fluid observed 
at RHIC; in particular they are not compatible with $\eta/s$ values extracted from viscous
fluid dynamics analysis \cite{Luzum:2008cw} of elliptic flow data. 
By increasing the coupling constant we find values of $\eta/s$ that are compatible with the RHIC data, but only for values of the coupling at which the perturbative
assumptions of the PCM may not anymore be valid. Inclusion of quantum-coherence effects,
such as the LPM effect, multi-particle scattering processes or 
turbulent color fields leading to an anomalous viscosity \cite{Asakawa:2006tc,Asakawa:2006jn} may
explain the origin of the observed small $\eta/s$ values, but the final determination
of this question remains to be settled in future work.

\section{Acknowledgments}
SAB would like to thank Peter Arnold, Carsten Greiner, Berndt M\"uller and Zhe Xu for 
helpful discussions.
This work was supported by the DOE under grant DE-FG02-05ER41367. JF acknowledges 
support from the TUNL REU program under NSF award NSF-PHY-08-51813.
The calculations were performed using resources provided by the Open Science Grid, 
which is supported by the National Science Foundation and the U.S. Department of 
Energy's Office of Science.

\section*{References}
%\begin{thebibliography}{10}
\bibliographystyle{iopart-num}
\bibliography{/Users/bass/Publications/SABrefs}
%\end{thebibliography}
\end{document}